\documentclass{iopart}
\usepackage[utf8]{inputenc}
\usepackage[T1]{fontenc}
\usepackage[greek,main=english]{babel}
\languageattribute{greek}{ancient}
\usepackage{csquotes}
\usepackage{alphabeta}
\usepackage{iopams}
\usepackage{upgreek}
\usepackage{amsthm}
\usepackage{hyperref}
\usepackage[%
	backend=biber,
	style=numeric,
	hyperref=true,
	sorting=none
]{biblatex}

\begin{document}

\title{James Clerk Maxwell on quantities and units}
\author{Michael P. Krystek}
\address{Physikalisch-Technische Bundesanstalt,\\ Bundesallee 100, D-38116 Braunschweig, Germany}
\eads{\mailto{Michael.Krystek@ptb.de}}

\begin{abstract}
	In the scientific literature the equation $Q=\{Q\}[Q]$ is frequently quoted, where $Q$ denotes a quantity, $\{Q\}$ a numerical value, and $[Q]$ a unit. During the last years some experts claimed, that this equation is due to James Clerk Maxwell. This is obvious, they say, from the opening sentences at the beginning of Maxwell's book \emph{A Treatise on Electricity and Magnetism}. In this paper it will be shown that this view cannot be justified.
\end{abstract}

\noindent{\it Keywords\/}: Maxwell, quantity, quantity value, magnitude, unit.

\section{Introduction}

In the scientific literature, especially in the context of metrology, a statement such as \emph{physical quantity $=$ pure number $\times$ unit} is frequently used, sometimes also expressed by the equation
\begin{equation*}
	Q=\{Q\}[Q]\,,
\end{equation*}
where $Q$ denotes a quantity, $\{Q\}$ a numerical value, and $[Q]$ a unit, both related to the quantity $Q$. In discussions during the last years some experts claimed, that this equation is due to James Clerk Maxwell. This is obvious, they say, from the opening sentences
\begin{quote}
	\emph{1.] EVERY expression of a Quantity consists of two factors or components. One of these is the name of a certain known quantity of the same kind as the quantity to be expressed, which is taken as a standard of reference. The other component is the number of times the standard is to be taken in order to make up the required quantity. The standard quantity is technically called the Unit, and the number is called the Numerical Value of the quantity.}
\end{quote}
at the beginning of Maxwell's book \emph{A Treatise on Electricity and Magnetism} \cite[p.~1]{Maxwell}. These experts also say that this quotation confirms their view that a unit is a quantity. In the following I will quote from Maxwell's treatise and make some comments to show that this view cannot be justified.

\section{Maxwell's understanding of the concepts ‘quantity’ and ‘unit’}

If we look more closely at the first paragraph of the first chapter of Maxwell's book \emph{A Treatise on Electricity and Magnetism} \cite[p.~1]{Maxwell}, we first notice that Maxwell did not define the term ‘quantity’. Maxwell just speaks of an “expression of a quantity” which “consists of two factors or components”. But this does not fulfil the requirements of a sound definition of the term ‘quantity’. As de Boer said \cite[p.~406]{deBoer}
\begin{quote}
	\emph{\dots Maxwell does not speak here of a definition of the concept “physical quantity”, but rather gives a factual description of this concept as it is actually used. This is important in connection with the current view that the concept quantity is a concept basic to the whole of quantity calculus, \dots .}
\end{quote}
The words “expression” and “factor” are interpreted by the supporters of the the aforementioned thesis as an indication that Maxwell meant the above mentioned equation. However, this is very doubtful for several reasons, as we shall see. Moreover, the question arises why a mathematically oriented scientist like Maxwell did not simply write down the equation in order to rule out any misunderstanding from the very beginning. The words “factors” or “components” can certainly not be understood to mean that Maxwell was speaking here of the operands of a multiplication, {i.\,e.}\ a binary mathematical operation. This is clearly contradicted by the fact that the next sentence says “One of these is the name of a certain known quantity”, and in the Table of Contents we find
\begin{quote}
	\emph{1. The expression of a quantity consists of two factors, the numerical value, and the name of the concrete unit \dots}
\end{quote}
This tells us that for Maxwell “a certain known quantity” was a “concrete unit”, {i.\,e.}\ as de Boer said \cite[p.~406]{deBoer}
\begin{quote}
	\emph{\dots the concept “unit” is described as referring to, or being the name of, a standard, the physical or material realization of the unit, also expressed by the French term étalon \dots}
\end{quote}
This view is supported by the quotations \cite[p.~2]{Maxwell}
\begin{quote}
	\emph{3.] (1) Length. The standard of length for scientific purposes in this country is one foot, which is the third part of the standard yard preserved in the Exchequer Chambers.}
\end{quote}
\begin{quote}
	\emph{4.] (2) Time. The standard unit of time in all civilized countries is deduced from the time of rotation of the earth about its axis. The sidereal day, or the true period of rotation of the earth, can be ascertained with great exactness by the ordinary observations of astronomers; and the mean solar day can be deduced from this by our knowledge of the length of the year.}
\end{quote}
\begin{quote}
	\emph{The unit of time adopted in all physical researches is one second of mean solar time.}
\end{quote}
and
\begin{quote}
	\emph{5.] (3) Mass. The standard unit of mass is in this country the avoirdupois pound preserved in the Exchequer Chambers. The grain, which is often used as a unit, is defined to be the 7000th part of this pound.}
\end{quote}
This clearly indicates that Maxwell's standard quantity -- “technically called the Unit”, as he said -- was an artefact (like the \emph{International Prototype of the Kilogram}, called the IPK, of the SI until recently), as it was common at his time. But an artefact or its name, {i.\,e.}\ a designation of an object, cannot serve as a factor of the mathematical operation multiplication.

What Maxwell wanted to express with the sentences of the first paragraph of his treatise might become more clear from the quotation \cite[p.~3]{Maxwell}
\begin{quote}
	\emph{\dots call the unit of length $[L]$. If $l$ is the numerical value of a length, it is understood to be expressed in terms of the concrete unit $[L]$, so that the actual length would be fully expressed by $1[L]$.}
\end{quote}
Here the expression $1[L]$ does not denote a product of the number one, $1$, and the concrete unit $[L]$, but rather means a \emph{denominate number} (“benannte Zahl”), as Helmholtz, a contemporary of Maxwell, has called it \cite[p.~375]{Helmholtz}
\begin{quote}
	\emph{Objecte oder Attribute von Objecten, die mit ähnlichen verglichen den Unterschied des grösser, gleich oder kleiner zulassen, nennen wir Grössen. Können wir sie durch eine benannte Zahl ausdrücken, so nennen wir diese den Werth der Grösse, \dots}
\end{quote}
\begin{quote}
	[Objects or attributes of objects, which compared with similar objects allow the distinction into greater, equal or smaller, we call magnitudes. If we can express them by a denominate number, we call the latter the value of the magnitude, \dots]
\end{quote}
This view is in principle confirmed by de Boer \cite[p.~407]{deBoer}, although his quotation of Helmholtz
\begin{quote}
	\emph{Objects or attributes of objects, which — compared with others of the same kind — permit the distinctions ‘larger’, ‘equal’ or ‘smaller’ are called ‘quantities’. If these can be expressed by a ‘concrete number’ then we call this the ‘value of the quantity’.}
\end{quote}
is not a proper quotation (it is an interpretation according to de Boer's intentions), as the reader may verify.

Maxwell said \cite[p.~1 and p.~2]{Maxwell}
\begin{quote}
	\emph{The formulae at which we arrive must be such that a person of any nation, by substituting for the different symbols the numerical value of the quantities as measured by his own national units, would arrive at a true result.}
\end{quote}
{i.\,e.}\ for Maxwell the symbols of his equations meant numerical values of the quantities. This interpretation was common at his time. Physicists used what Julius Wallot called “Zahlenwertgleichungen” \cite[p.~29]{Wallot1926}, and consequently needed a dimensional analysis, in order to check the correctness of their equations.

Maxwell said \cite[p.~6]{Maxwell}
\begin{quote}
	\emph{In transforming the values of physical quantities determined in terms of one unit, so as to express them in terms of any other unit of the same kind, we have only to remember that every expression for the quantity consists of two factors, the unit and the numerical part which expresses how often the unit is to be taken. Hence the numerical part of the expression varies inversely as the magnitude of the unit, that is, inversely as the various powers of the fundamental units which are indicated by the dimensions of the derived unit.}
\end{quote}
This text is remarkable, because it shows that Maxwell was aware of the fact that every quantity must be assigned a unit of the same kind (which was introduced by him as a material standard), in order to apply mathematical operations. The “expression of a Quantity” was thus for him a product of a number (the “numerical part”) and “the magnitude of the unit” (which Helmholtz called “benannte Zahl”, {i.\,e.}\ a \emph{denominate number}). But then, according to the rules of mathematics, this product (the “expression of a Quantity”) is itself a \emph{denominate number} as well.

On \cite[p.~251]{Maxwell} we read:
\begin{quote}
	\emph{This investigation is approximate only when b1 and b2 are large compared with a, and when a is large compared with c. The quantity a is a line which may be of any magnitude. It becomes infinite when c is indefinitely diminished.}
\end{quote}
This text shows, that for Maxwell a line was a quantity and its length was a magnitude. Compare this quotation with the statement of B. Russell \cite[p.~159]{Russell}
\begin{quote}
	\emph{An actual footrule is a quantity : its length is a magnitude}
\end{quote}
It turns out, then, that it is necessary to read Maxwell's entire treatise, and not just the opening sentences, to understand what the sentences at the beginning of the first chapter really mean. I contend that these sentences do not, as claimed, support the definitions of the terms ‘quantity’, ‘unit’ and ‘quantity value’ as given in the current version of the \emph{International vocabulary of metrology (VIM)} \cite{VIM}. For Maxwell, a unit was just an artefact, like until recently the \emph{International Prototype of the Kilogram} in the SI.

\section{Maxwell and the C.G.S. System of Units}

Finally, let's take a look at another piece of information in order to clarify what was meant by the term ‘unit’ in Maxwell's times. It is probably little known that Maxwell was a member of the \emph{Committee for the Selection and Nomenclature of Dynamical and Electrical Units}. In the year 1875 (two year after the publication of Maxwell’s treatise), the Secretary of this committee, Prof. J. D. Everett, published a book on behalf of the committee. I quote from the second edition of this book, published in 1979 (the year of Maxwell's untimely death) \cite{Everett}.

The committee members stated in their second report (added as Appendix to Prof. Everett’s book) that
\begin{quote}
	\emph{They believe, however, that, in order to render their recommendations fully available for science teaching and scientific work, a full and popular exposition of the whole subject of physical units is necessary, together with a collection of examples (tabular and otherwise) illustrating the application of systematic units to a variety of physical measurements. Students usually find peculiar difficulty in questions relating to units ; and even the experienced scientific calculator is glad to have before him concrete examples with which to compare his own results, as a security against misapprehension or mistake.}
\end{quote}
\begin{quote}
	\emph{Some members of the Committee have been preparing a small volume of illustrations of the C.G.S. system [Centimetre-Gramme-Second system] intended to meet this want.}
\end{quote}
Prof Everett added the remark
\begin{quote}
	\emph{[The first edition of the present work is the volume of illustrations here referred to].}
\end{quote}
In the Preface to the first edition (reprinted on page V of the second edition) we read
\begin{quote}
	\emph{I am indebted to several friends for assistance in special departments but especially to Professor Clerk Maxwell and Professor G. C. Foster, who revised the entire manuscript of the work in its original form.}
\end{quote}
This ensures that Maxwell agreed with the content of Everett's book and also with the following statement on page 1, dealing with the General Theory of Units:
\begin{quote}
	\emph{1. THE numerical value of a concrete quantity is its ratio to a selected magnitude of the same kind, called the unit.}
\end{quote}
\begin{quote}
	\emph{Thus, if $L$ denote a definite length, and $l$ the unit length, $L/l$ is a ratio in the strict Euclidean sense, and is called the numerical value of $L$.}
\end{quote}
\begin{quote}
	\emph{The numerical value of a concrete quantity varies directly as the concrete quantity itself, and inversely as the unit in terms of which it is expressed.}
\end{quote}
Here the remark “\dots in the strict Euclidean sense, \dots” undoubtedly refers to the definition of the term magnitude (μέγεθος) which appears in the literature for the first time in Book V of Euclid's elements \cite[p.~113]{Euclid}.

Thus the term ‘magnitude’ was understood by the scientists of Maxwell's and Helmholtz's times in the sense of Euclid's definitions and a ‘unit’ was only a particular magnitude of the same kind as the magnitude attributed to the quantity under consideration. Just as B. Russell wrote: “An actual footrule is a quantity : its length is a magnitude”, {i.\,e.}\ for any length (a magnitude, not a quantity, since for Maxwell a \emph{line} was a quantity \cite[p.~251]{Maxwell}) the unit was the magnitude of an artefact, namely the footrule, {i.\,e.}\ “The standard of length for scientific purposes in this country is one foot, which is the third part of the standard yard preserved in the Exchequer Chambers”, as Maxwell said.

\section{Conclusion}

In summary, we must conclude that we cannot rely on Maxwell's treatise when defining important metrological terms such as ‘quantity’ or ‘unit’ for the purposes of the revised SI. The unit of Maxwell and his contemporaries was a material standard and not the abstract concept of a unit underlying modern thinking based on what is now called the “Quantity Calculus”. As de Boer stated unequivocally \cite[p.~412]{deBoer}
\begin{quote}
	\emph{In the past and even today many scientists use the name “unit” for what now should properly be called a “standard” for a unit. A standard for the unit of a particular quantity is realized by a real physical system for which the numerical value of the quantity concerned is equal to 1.}
\end{quote}
In conclusion, it should be noted that the SI of today is a purely abstract system which needs a solid logical and mathematical foundation. This cannot be achieved with a terminology that tries to preserve the thinking of the past.

\printbibliography

@Book{Maxwell,
	author = {Maxwell~J.~Clerk},
	title = {A Treatise on Electricity and Magnetism},
	volume = {I},
	publisher = {Clarendon Press},
	year = {1873},
	address = {Oxford},
	language = {english},
}

@article{deBoer,
	author = {de Boer~J.},
	title = {On the history of quantity calculus and the international system},
	journal = {Metrologia},
	year = {1995},
	number = {31},
	pages = {405--429},
	language = {english},
}

@inbook{Helmholtz,
	author = {Helmholtz~H.},
	title = {Zählen und Messen, erkenntnisstheoretisch betrachtet},
	booktitle = {Philosophische Aufsätze},
	booktitleaddon = {Eduard Zeller zu seinem fünfzigjährigen Doctorjubiläum gewidmet},
	publisher = {Fues' Verlag},
	year = {1887},
	address = {Leipzig},
	pages = {17--52},
	language = {german},
}

@Book{Euclid,
	author = {Euclid},
	title = {The Thirteen Books of Euclid’s Elements; translated by T.~L.~Heath},
	publisher = {University Press},
	year = {1968},
	address = {Cambridge},
}

@Book{Russell,
	author = {Russell.~B.},
	title = {The Principles of mathematics},
	publisher = {W. W. Norton and Company Inc.},
	year = {1937},
	address = {New York},
	language = {english},
}

@Book{Everett,
	author = {Everett.~J.~D.},
	title = {Units and physical constants},
	publisher = {Macmillan and Co.},
	year = {1879},
	address = {London},
	language = {english},
}

@inbook{Wallot1926,
	author = {Wallot~J.},
	title = {Dimensionen, Einheiten, Maßsysteme},
	booktitle = {Handbuch der Physik},
	volume = {II},
	booktitleaddon = {Elementare Einheiten und ihre Messung},
	editor = {Geiger~H. and Scheel~K.},
	publisher = {Julius Springer Verlag},
	year = {1926},
	address = {Berlin},
	pages = {1--41},
	language = {german},
}

@book{VIM,
	author = {{ISO/IEC} {G}uide 99},
	title = {{I}nternational vocabulary of metrology --- {B}asic and general concepts and associated terms ({VIM})},
	publisher = {{I}nternational {O}rganization for {S}tandardization ({ISO})},
	year = {2007},
	address = {Gen\`eve},
	language = {english}
}

\end{document}